\documentclass[preprint2]{aastex}
\usepackage{epsfig}
\usepackage{rotating}
\usepackage{natbib}
\usepackage{float}
\usepackage{multirow}
\usepackage{comment}
\usepackage{threeparttable}
\newcommand{\msun}{{\rm M}_{\sun}} 
\shorttitle{Nonthermal electrons in hot accretion flows}
\shortauthors{Nied\'zwiecki et al.}
\begin{document} 
\title{On the role and origin of nonthermal electrons \\
 in hot accretion flows}   
\author{Andrzej Nied\'zwiecki, Agnieszka St\c epnik}
\affil{Department of Astrophysics, University of \L \'od\'z, Pomorska 149/153, 
90-236 \L \'od\'z, Poland}
\email{niedzwiecki@uni.lodz.pl, agajer@o2.pl}
\author{Fu-Guo Xie}
\affil{Key Laboratory for Research in Galaxies and Cosmology, Shanghai
  Astronomical Observatory, Chinese Academy of Sciences,
80 Nandan Road, Shanghai 200030, China}   
\email{fgxie@shao.ac.cn}

\begin{abstract}
We study the X-ray spectra of tenuous, two-temperature accretion flows using a
model involving an exact, Monte Carlo computation of the global
Comptonization effect as well as general relativistic description of
both the flow structure and radiative processes. In our previous work
we found that in flows surrounding supermassive black holes, thermal
synchrotron radiation is not capable of providing a sufficient seed
photons flux to explain the X-ray spectral indices as well as the
cut-off energies measured in several best-studied AGNs.  In this work
we complete the model by including seed photons provided by nonthermal
synchrotron radiation and we find that it allows to reconcile the hot
flow model with the AGN data.  We take into account two possible
sources of nonthermal electrons.  First, we consider $e^{\pm}$
produced by charged-pions decay, which should be always present in the
innermost part of a two-temperature flow due to proton-proton
interactions. We find that for a weak heating of thermal electrons
(small $\delta$) the synchrotron emission of pion-decay $e^{\pm}$ is
much stronger than the thermal synchrotron emission in the considered
range of bolometric luminosities, $L \sim (10^{-4}-10^{-2}) L_{\rm
  Edd}$.  The small-$\delta$ model including hadronic effects in
general agrees with the AGN data, except for the case of a slowly
rotating black hole and a thermal distribution of protons. For large
$\delta$, the pion-decay $e^{\pm}$ have a negligible effect and then
in this model we consider nonthermal electrons produced by direct
acceleration. We find an approximate agreement with the AGN data for
the fraction of the heating power of electrons which is used for the
nonthermal acceleration $\eta \sim 0.1$. However, for constant $\eta$
and $\delta$, the model predicts a positive correlation of the X-ray
spectral index with the Eddington ratio, and hence a fine tuning of
$\eta$ and/or $\delta$ with the accretion rate is required to explain the
negative correlation observed at low luminosities. We note a significant
difference between the dependence of plasma parameters, $T_{\rm e}$ and $\tau$, 
on the Eddington ratio that is predicted by the large- and small-$\delta$ models.
This may be the key property allowing for estimation
of the value of $\delta$. However, a precise measurement of the spectral
cut-off is required and we note that differences between results
available in literature are similar in magnitude to the difference
between the model predictions.  In flows surrounding stellar-mass
black holes, the synchrotron emission of pion-decay $e^{\pm}$ exceeds
the thermal synchrotron only above $\sim 0.01 L_{\rm Edd}$.
Furthermore, in such flows the nonthermal synchrotron radiation is
emitted at energies $\ga 1$ keV, and therefore the Compton cooling is
less efficient than in flows surrounding supermassive black holes.
This may explain spectral differences between AGNs and black-hole
transients around $\sim 0.01$ $L_{\rm Edd}$ (the latter being
typically much harder).
\end{abstract}
 
\section{Introduction} 
\label{intro} 

Optically thin, hot accretion flows are widely considered as a
relevant accretion mode below $\sim 0.01 L_{\rm Edd}$ in black-hole
binaries as well as  in AGNs, and the X-ray radiation - typically
dominating  the radiative output at low luminosities - is most likely
produced by thermal Comptonization in the inner parts of such flows
(see reviews in Zdziarski \& Gierli\'nski 2004, Done et al.\ 2007,
Yuan \& Narayan 2014, Poutanen \& Veledina 2014).  In our previous
work (Nied\'zwiecki et al.\ 2014, hereafter N14) we considered the
standard version of hot-flow models, with seed photons for
Comptonization provided mainly by thermal synchrotron radiation. We
found that such a model roughly agrees with observations of several
well studied black hole transients for bolometric luminosities between
$\sim 10^{-4}  L_{\rm Edd}$ and $10^{-2} L_{\rm Edd}$. For a small
sample of well studied AGNs, observed in the same luminosity range, we
found a strong disagreement with the model predictions; the predicted
spectra are much harder than observed and their cut-off energies are
too high.

This clearly indicates that in hot flows around supermassive black
holes, thermal synchrotron emission is not capable of providing the
required flux of seed photons for thermal Comptonization, which is the
main cooling process at such luminosities.  In this paper we consider
extension of the model, taking into account the presence of
nonthermal electrons and their nonthermal synchrotron emission. The
presence of such a nonthermal component can significantly increase the
efficiency of Compton cooling (see Wardzi\'nski \& Zdziarski 2001,
Veledina et al.\ 2011);  while thermal synchrotron is strongly
self-absorbed, the non-thermal synchrotron is emitted at higher
frequencies and subject to weaker self-absorption, therefore, it is
a much more efficient source of seed photons.  See also Poutanen \&
Veledina (2014) for arguments supporting the presence of nonthermal
electrons in hot flows.

Our model essentially follows the original formulation of advection
dominated accretion flow (ADAF) model (e.g.\ Narayan \& Yi 1995).
Relatively weak Coulomb coupling between protons and electrons in a
tenuous plasma results in a two-temperature structure, which is a key
property  of ADAF solutions, as such flows are supported by proton
pressure.  In the innermost part of the flow, the hot protons have
energies above the threshold for pion production.  Therefore,
relativistic $e^{\pm}$ from charged-pions decay should be always
present in ADAFs.  Their nonthermal synchrotron emission was
previously studied by Mahadevan (1999, hereafter M99) for the radio
emission of Sgr A*, however, it has never been self-consistently
implemented as a source of seed photons in models of high-energy
emission.

Nonthermal particles may be also directly produced by nonthermal
acceleration, e.g.\ by magnetic reconnection induced by the
magneto-rotational instability (cf.\ Riquelme et al.\ 2012). We take
it into account by considering both a nonthermal distribution of
protons in hadronic models and the presence of a nonthermal population
of electrons not related with the $\pi^{\pm}$-decay rate. In the latter case 
the amount of nonthermal electrons is a free parameter (in contrast to
the hadronic model); such models with simple descriptions of accretion flow
(one-zone or approximate radial dependence) have been considered in the context
 of ADAF solutions e.g.\ by Malzac \& Belmont (2009), Veledina et al.\ (2011).

\section{Model}
\label{sec:flow}

We apply the model developed in our recent works, see Nied\'zwiecki et
al.\ (2012, hereafter N12) and N14, except for the source of seed
photons, which includes nonthermal synchrotron radiation, whereas in
the previous studies we took into account only thermal synchrotron
emission (and bremsstrahlung, which is negligible for accretion rates
considered in our works).  We consider a black hole, characterized by
its mass, $M$, and angular momentum, $J$, surrounded by a
geometrically thick accretion flow with an accretion rate, $\dot
M$. We define the following dimensionless parameters: $r = R / R_{\rm
  g}$, $a = J / (c R_{\rm g} M)$, $\dot m = \dot M / \dot M_{\rm Edd}$,
where $\dot M_{\rm Edd}= L_{\rm Edd}/c^2$, $R_{\rm g}=GM/c^2$ is the
gravitational radius and $L_{\rm Edd} \equiv 4\pi GM m_{\rm p}
c/\sigma_{\rm T}$ is the Eddington luminosity.  We assume that  the
density distribution is given by $n(R,z)=n(R,0)
\exp(-z^2/2H^2)$, where $H$ is the height scale at $R$ and $z=R \cos
\theta$. We define the vertical optical depth as $\tau = H \sigma_{\rm T} n$.

We assume the viscosity parameter, $\alpha=0.3$, and the
ratio of the gas pressure (electron and ion) to the magnetic pressure,
$\beta$. The fraction of the dissipated energy which heats directly
electrons is denoted by $\delta$.

We find the global hydrodynamical solution of the general relativistic
(GR) structure  equations following Manmoto (2000) with modifications
described in N12.  Most importantly, we use the global Compton cooling
rate computed using a GR Monte Carlo (MC) method, whereas in similar
studies local approximations are usually applied (see Yuan et
al.\ 2009, Xie et al.\ 2010, N12 and N14 for discussion of related
inaccuracies). We find the self-consistent electron temperature
distribution, $T_{\rm e}(r)$, by iterating between the solutions of the electron energy
equation  and the GR MC Comptonization simulations until we find
mutually consistent solutions. This procedure involves assumption that
other parameters of the flow (density, proton temperature, height
scale, velocity field) are not affected by changes in $T_{\rm e}$,
which limits the maximum luminosity of the flow to $\sim 0.01 L_{\rm
  Edd}$ (at larger luminosities the Coulomb cooling of protons becomes
important and the flow is characterized by a dramatic dependence on
even small changes of $T_{\rm e}$). The minimum luminosity of flows
which can be studied with our current model, $\sim 10^{-4}
L_{\rm Edd}$, is given by the requirement that the energy balance for
electrons is determined by radiative cooling (rather than advection).

The total luminosity detected far away from the flow is denoted by $L$ and
the luminosity detected in the 2-10 keV range by $L_{2-10}$.

The thermalization as well as cooling time-scales for protons are much
longer than the accretion time-scale, $\simeq R/|v^r|$, so the
distribution of protons is determined by heating/acceleration
processes which are poorly understood. We then consider two limiting cases by assuming that (i)
all protons have a Maxwellian distribution; (ii) a small fraction of
protons has a power-law distribution, $n_{\rm pl}(\gamma) \propto
\gamma^{-s_{\rm p}}$, and the remaining protons are cold. These two models are
identical to models T and N, respectively, in Nied\'zwiecki et
al.\ (2013, hereafter N13); in particular, the fraction of protons
with the power-law distribution is given by equation (7) in N13. We also
consider (only in Fig.\ \ref{fig:agn}b) a hybrid model, with half of
the energy stored in the thermal population of protons and another half 
in the nonthermal (power-law) population. For
each distribution function of protons, their energy density at each
$r$ is equal to that obtained from the hydrodynamic solution. Note
that in the hydrodynamic solutions we always assume a thermal
distribution, however, for a nonthermal distribution the flow
structure should not change significantly (cf.\ Kimura et al.\ 2014). 

In the computation of the rate of $e^{\pm}$ production by charged-pion
decay and their nonthermal synchrotron radiation we follow M99, except
for interactions between power-law protons which are taken into
account in our model and neglected by M99; the simplification applied
by M99 underestimates the power in proton-proton interaction products
by a factor of $\sim 2$ (see N13). We also used the {\sc galprop} code
(Moskalenko \& Strong 1998) to test our computations of the production
and decay of $\pi^{\pm}$.  

For all distributions of protons, 
the energy spectrum of the produced $e^{\pm}$ has a
maximum at $\sim 35$ MeV. For a thermal distribution, it
declines exponentially at higher energies. For a power-law proton
distribution,  it is the power-law at high energies with the same slope
as the proton distribution, i.e.\ the injection index of electrons,
 $s_{\rm inj}$, equals $s_{\rm p}$.
For the range of parameters considered in this
work, at $r<1000$ for all relevant electron energies, the synchrotron
cooling time is much shorter than the accretion time, so in all
hadronic models the steady state electron distribution is $N(\gamma)
\propto \gamma^{-2}$ for Lorentz factors $\gamma \la 70$. 
 For $\gamma \gg 70$, $N(\gamma) \propto \gamma^{-(s_{\rm p}+1)}$ for the power-law 
distribution of protons, whereas $N(\gamma)$ decreases exponentially for the thermal 
distribution of protons (cf.\ figure 1 in M99).
We have also
checked that the Compton cooling time of nonthermal electrons is
longer than the synchrotron cooling time, so the steady state
distribution is determined by the synchrotron cooling rate.

The time-scale to establish pair equilibrium (with $e^{\pm}$ creation by pion decay
balanced by pair annihilation) is of the order of the accretion time scale (and much longer 
than the cooling time scale of nonthermal  $e^{\pm}$). Therefore, an approximate pair equilibrium 
should be established in the flow.
In all models the equilibrium density of pion-decay $e^{\pm}$ is at least by 
2 orders of magnitude lower than the density of the ionization electrons, 
so the hadronic $e^{\pm}$ contribute negligibly to the optical depth.

In models considering directly accelerated electrons we assume that a
fraction $\eta$ of the total power heating electrons, $\delta Q_{\rm
  diss}$ (where $Q_{\rm diss}(r)$ is the total power dissipated at
$r$), is used for direct acceleration of electrons, so at each radius
the power used for the acceleration is  $\eta \delta Q_{\rm diss}$,
while $(1 - \eta)  \delta Q_{\rm diss}$ is used for heating of thermal
electrons. 

In summary, we consider several versions of the hot-flow model.  The
purely thermal version, for which the thermal synchrotron radiation is
the only source of seed photons, is referred to as the {\it standard}
model {\bf S}. The extension of the model, taking into account
additional seed photons from nonthermal synchrotron of relativistic
$e^\pm$ produced by $\pi^\pm$ -decay, is referred to as the {\it
  hadronic model} and depending on the assumed distribution function
of protons it is denoted by {\bf HT} (for thermal protons), {\bf HN}
(for power-law proton distribution) or {\bf HH} (for hybrid
distribution with equal energies in a thermal and power-law proton
components). The model taking into account a {\it direct acceleration}
of electrons is referred to as model {\bf DA}. We consider model DA
only for $\delta=0.5$, for which hadronic processes have a negligible
effect (see below), so we do not consider cases mixing strong effects
of pion-decay and directly accelerated electrons. All results presented in this work
for model HN or HH correspond to the proton power-law index $s_{\rm p}=2.6$.

\begin{figure} 
\centerline{\includegraphics[width=8cm]{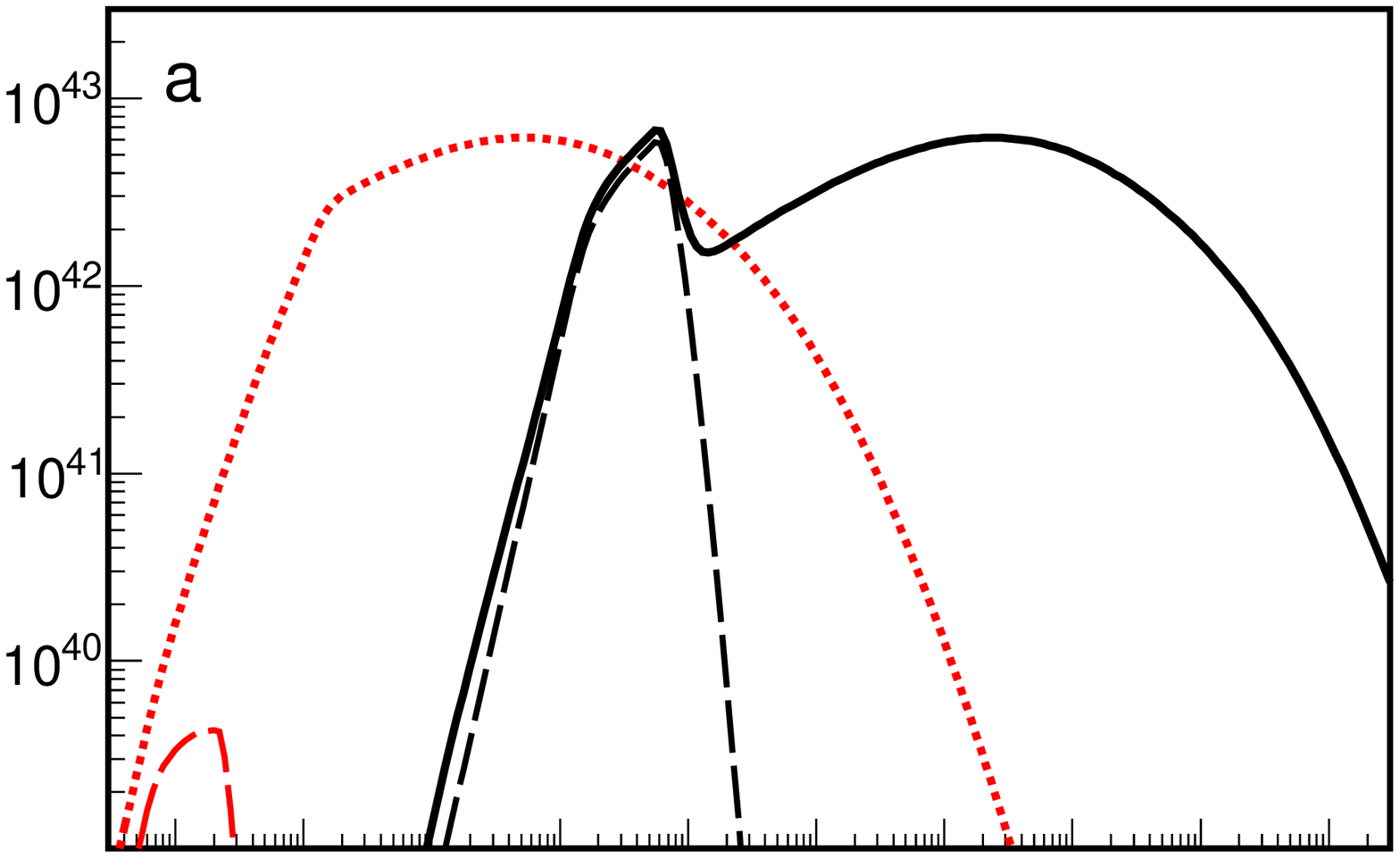}}
\centerline{\includegraphics[width=8cm]{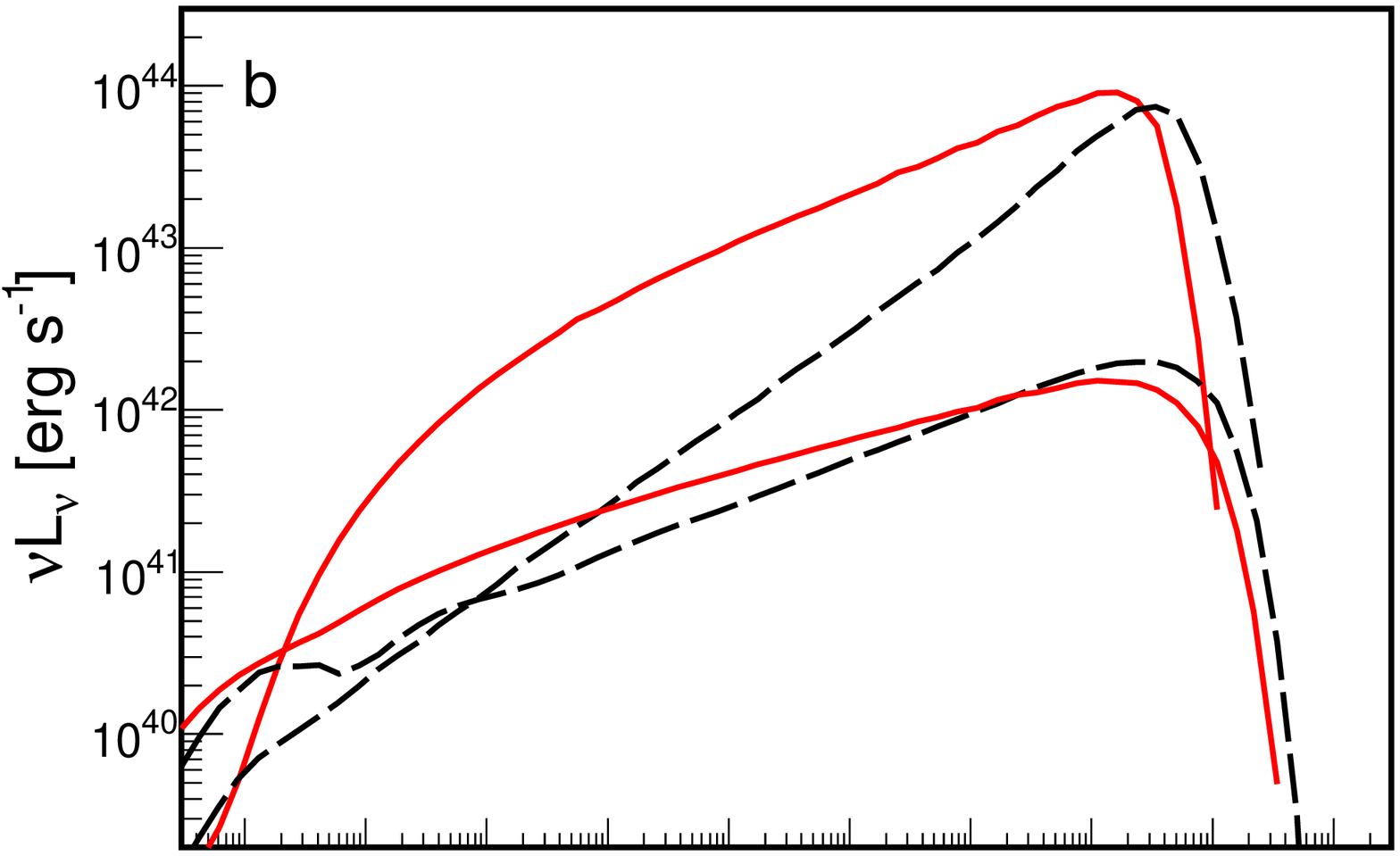}}
\centerline{\includegraphics[width=8cm]{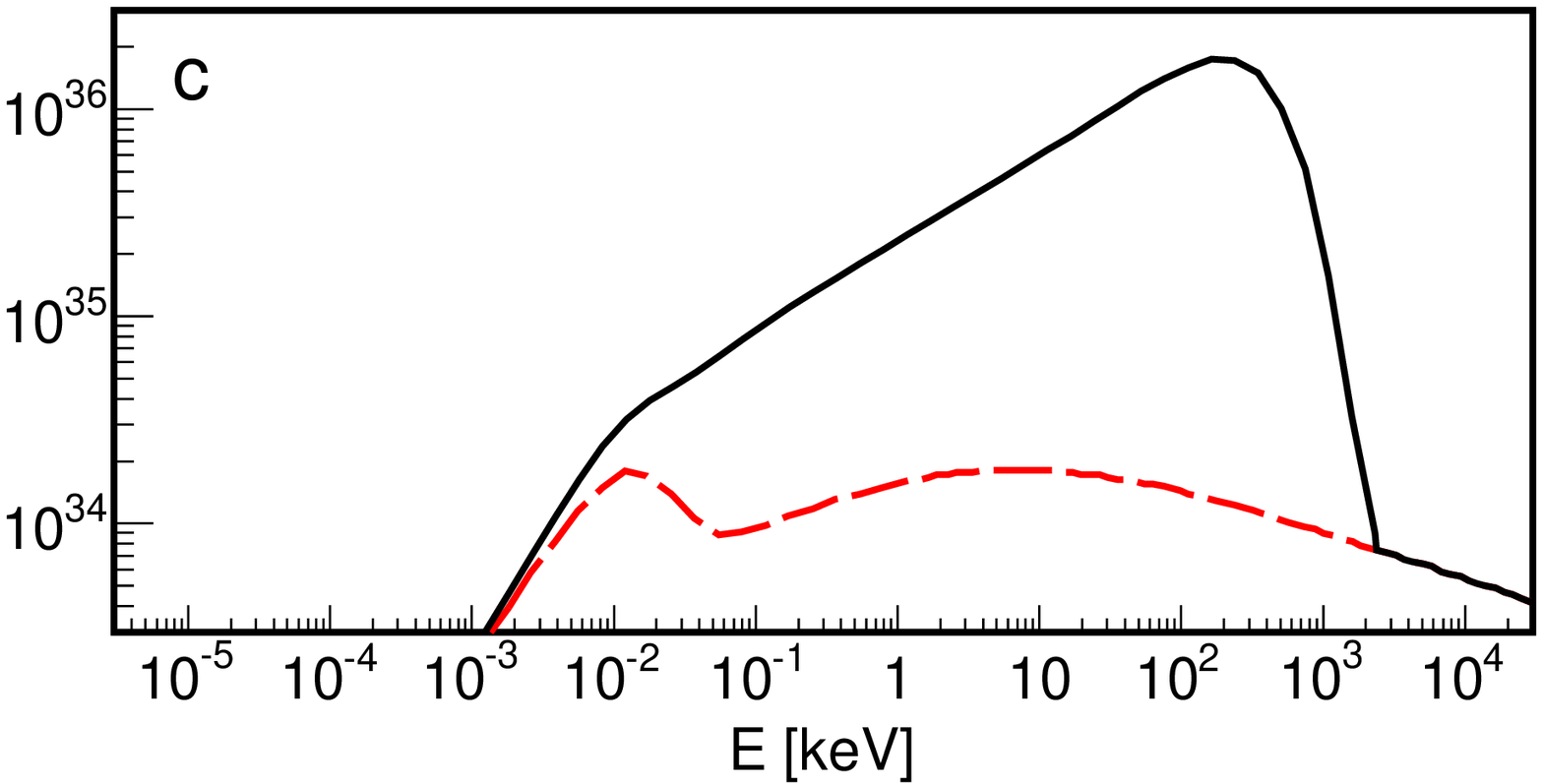}}
\caption{(a) {\it Rest-frame} spectra of the total synchrotron radiation for
  model HT with $a=0.95$, $\beta=1$, $\delta =10^{-3}$ and $\dot m =
  0.5$ are shown by the (red) dotted line for $M= 2 \times 10^8\,
  \msun$  and by the (black) solid line (scaled by $2 \times 10^7$)
  for $M= 10\, \msun$. Dashed lines show contribution from thermal
  synchrotron radiation. (b) (Red) solid lines show the {\it observed}
  spectra of synchrotron radiation and its thermal Comptonization for
  model HT with $a=0.95$, $\beta=1$, $\delta =10^{-3}$, $M= 2 \times
  10^8\, \msun$; the (black) dashed lines are for model S with the
  same parameters. In both models $\dot m = 0.1$ and 0.5 from bottom
  to top. (c) (Black) solid line shows the {\it observed} spectrum for
model HN  with $a=0.95$, $\beta=1$, $\delta =10^{-3}$, $M= 10\, \msun$ and
$\dot m = 0.5$;
(red) dashed line shows the contribution of synchrotron radiation.
}
\label{fig:spec} 
\end{figure}

\section{Results}

\subsection{Pion-decay electrons}
\label{sect:pion}

\begin{figure*} 
\centerline{\includegraphics[width=7cm]{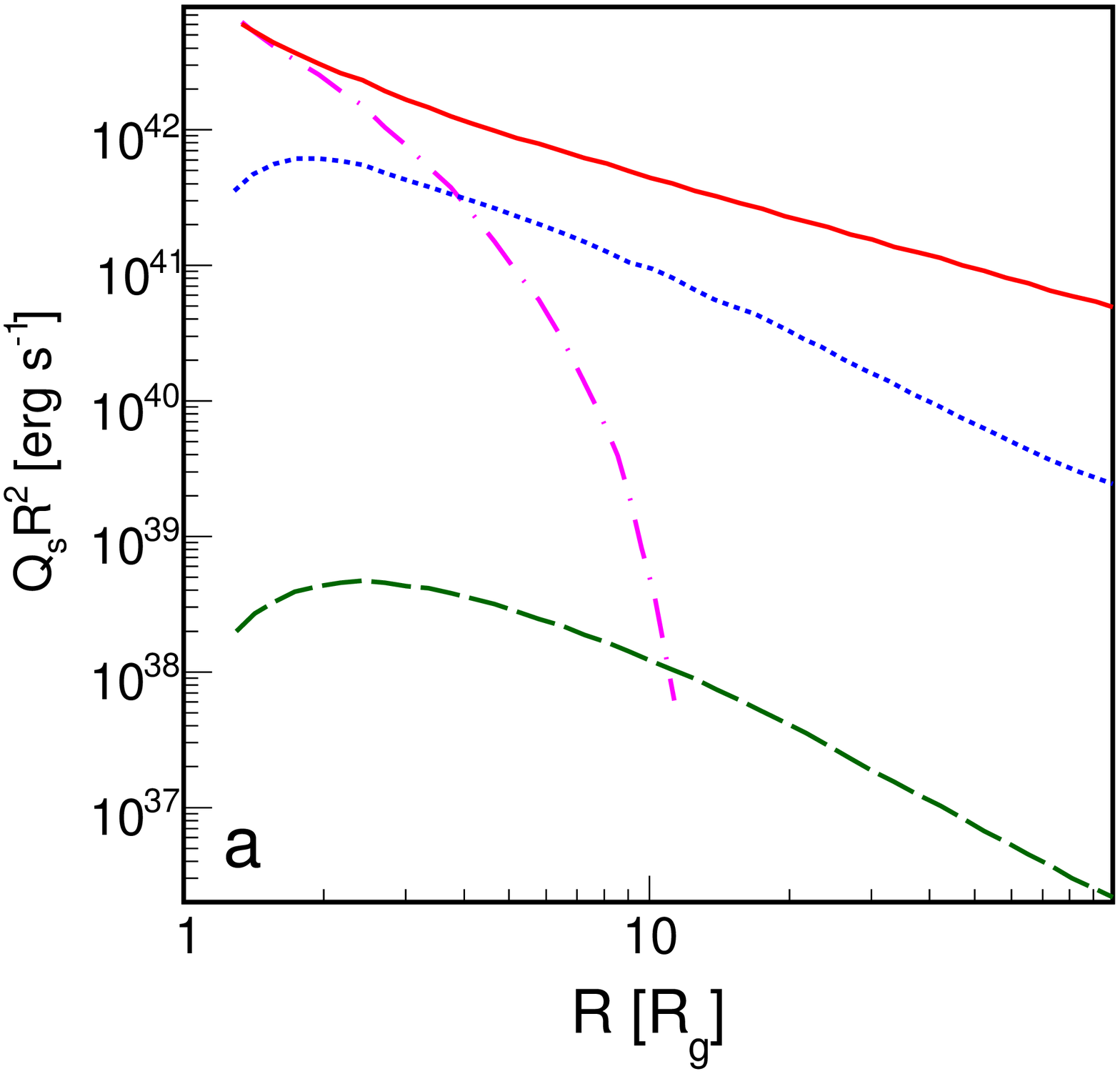} \includegraphics[width=7cm]{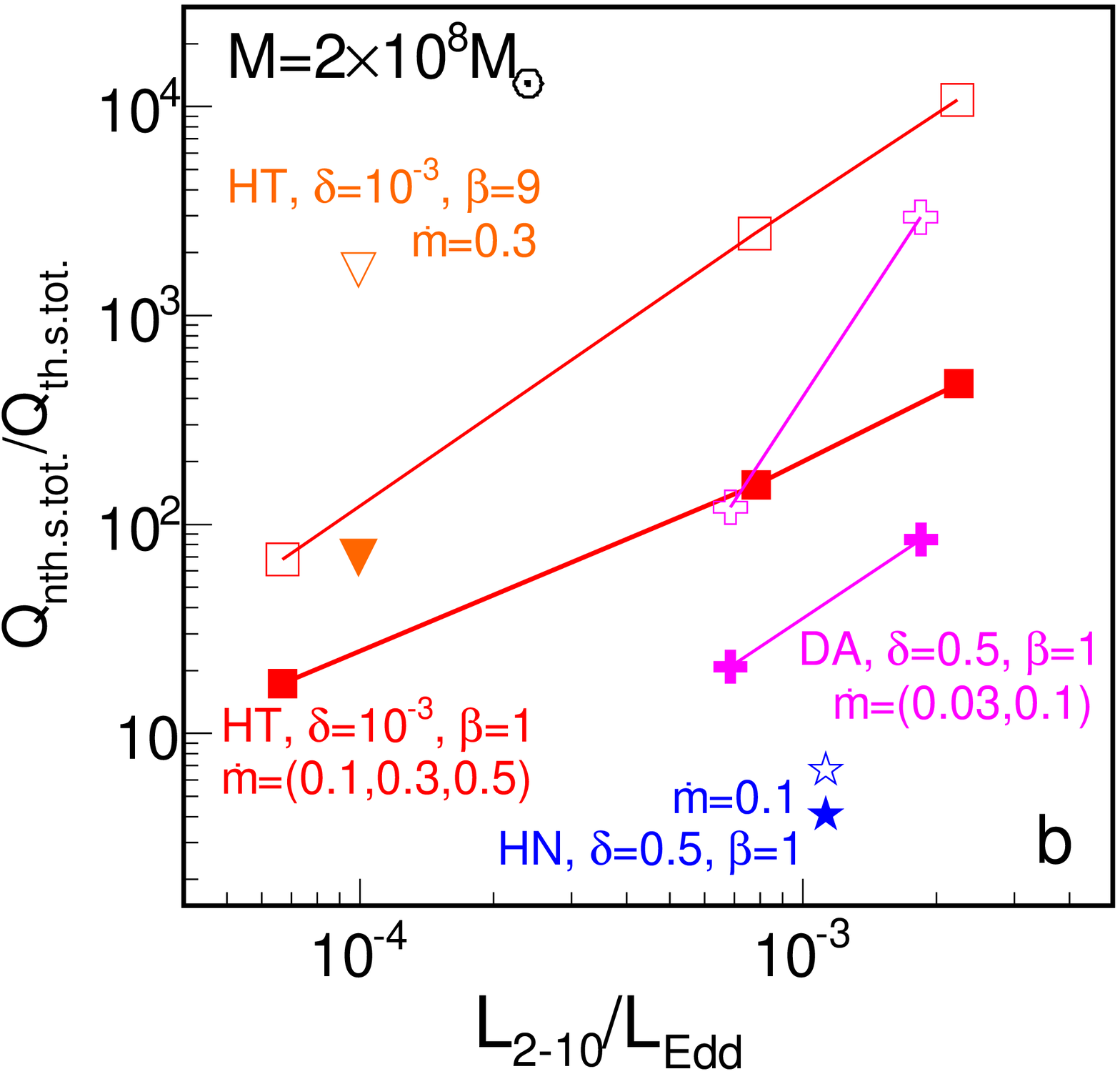}} 
\centerline{\includegraphics[width=7cm]{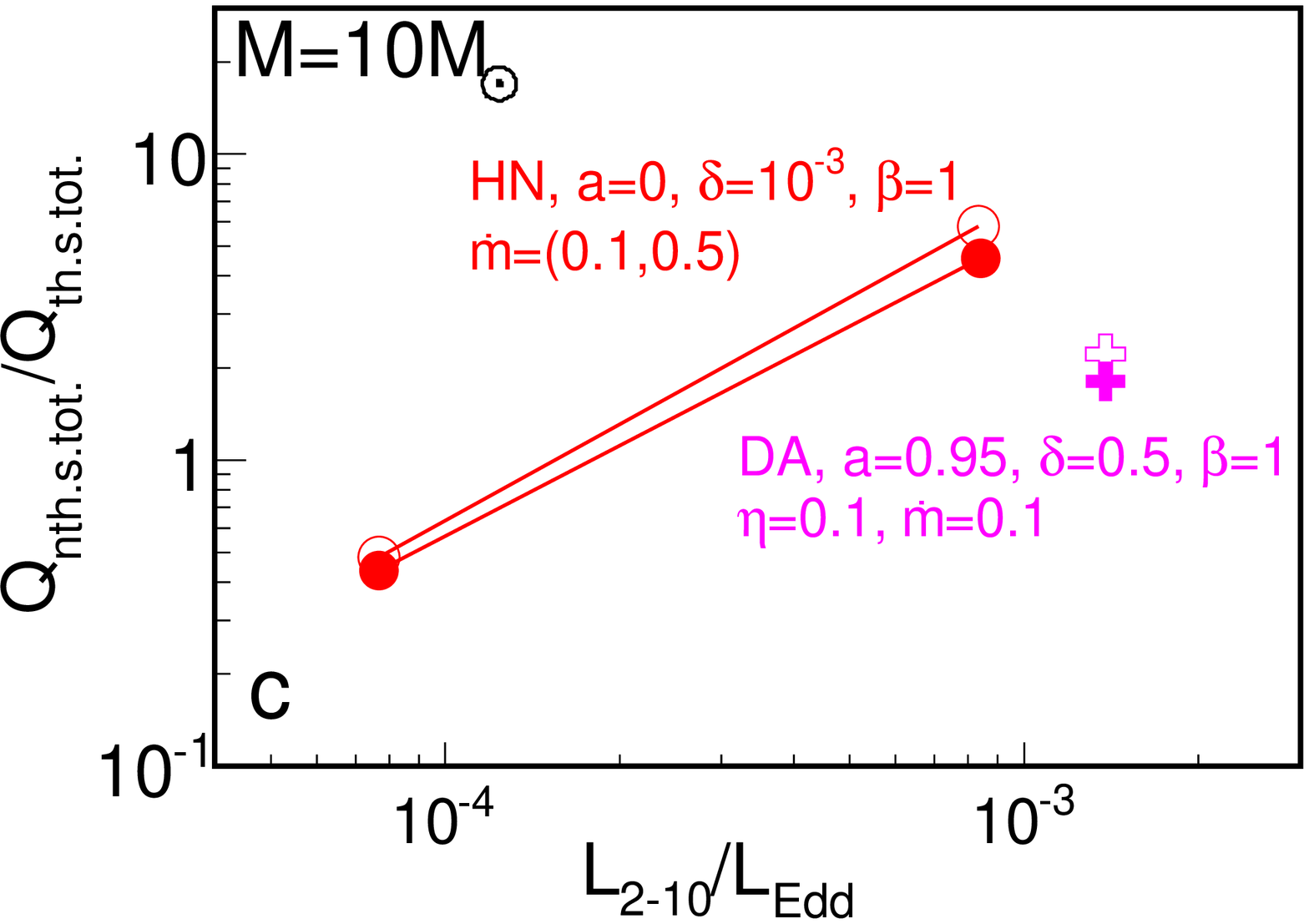} \includegraphics[width=7cm]{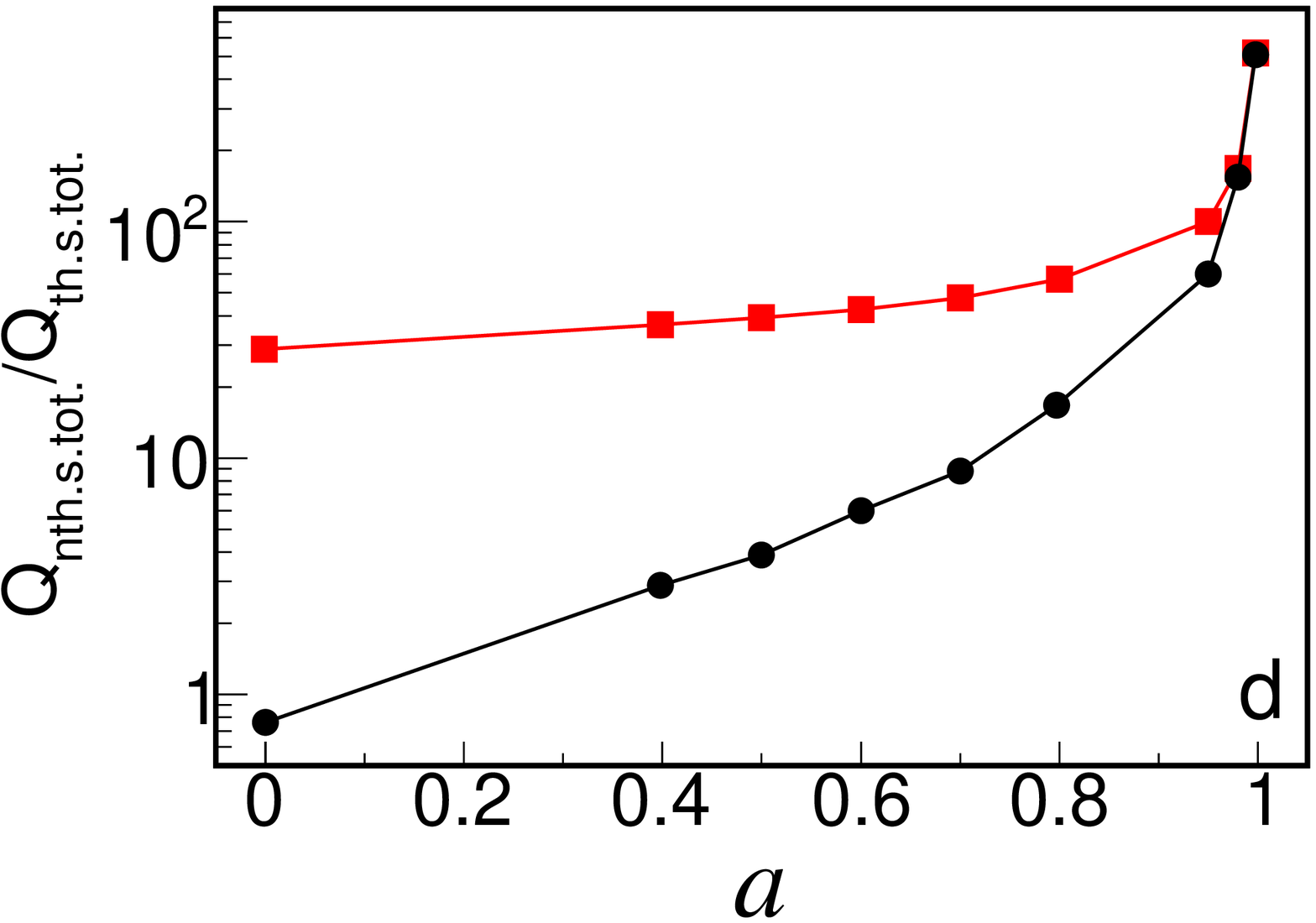} }
\caption{(a) Rest-frame, radial emissivity profiles for self-absorbed
  synchrotron radiation in models with $a=0.95$, $\beta=1$, $\delta
  =10^{-3}$ and $\dot m = 0.5$; all $Q_{\rm s}$ rates  are vertically
  integrated, so $Q_{\rm s}R^2$ gives the emissivity per unit volume
  times volume. (Blue) dotted and (green) dashed lines are for the
  thermal synchrotron radiation in model HT with $M= 10 \msun$ (scaled
  by $2 \times 10^7$)  and  $M= 2 \times 10^8 \msun$,
  respectively. (Red) solid and magenta (dot-dashed) lines are for the
  nonthermal synchrotron radiation in model HN and HT, respectively,
  with  $M= 2 \times 10^8\, \msun$ (or $M= 10\, \msun$  with the above
  scaling). In panels (b) and (c), open symbols show the ratio of $Q_{\rm nth.s.tot}$ to
  $Q_{\rm th.s.tot}$, as a function of the 2-10 keV Eddington ratio,
  in model DA (magenta crosses) and hadronic models (other symbols); full
  symbols show the ratio of the same $Q_{\rm nth,s,tot}$ to $Q_{\rm
    th.s.tot}$ in corresponding models S. Panel (b) is for $M= 2 \times 10^8\, \msun$; (red) squares: model HT  with $a=0.98$, (orange) triangle-down: model  HT with $a=0.95$;
(blue) star: model  HN with $a=0.95$; (magenta) crosses: model  DA with $\eta=0.1$ and $a=0.95$; other model parameters as specified in the figure.
Panel (c) is for $M= 10\, \msun$; (red) circles: model  HT,
(magenta) crosses: model DA (with $\gamma_0 = 20$).
(d) The ratio of $Q_{\rm nth.s.tot}$ in model HN (red squares), or HT (black
  circles), to $Q_{\rm th.s.tot}$ in corresponding model S, as a
  function of the black hole spin parameter, for $M= 2 \times 10^8
  \msun$, $\beta=1$, $\delta =10^{-3}$ and $\dot m = 0.3$.  }
\label{fig:synch} 
\end{figure*}

Figs \ref{fig:spec} and  \ref{fig:synch}  illustrate crucial
properties of the nonthermal synchrotron radiation produced by
pion-decay $e^{\pm}$, its effect on the Comptonized radiation and the
dependence on key parameters. Fig.\ \ref{fig:spec}a shows the spectral
distribution of the synchrotron radiation, for $M= 10\, \msun$  and
$M= 2 \times 10^8\, \msun$. For these values of $M$, the magnetic
field in the innermost part of the flow is $B \sim (10^6-10^8)$ G and
$B \sim (10^2-10^4)$ G, respectively. In all hadronic models,
regardless of the model parameters, the dominating contribution to the
nonthermal synchrotron component is produced by electrons with
$\gamma_0 \sim 100$ and the nonthermal synchrotron spectrum has a
maximum at $\nu_{\rm max}$ ($ \simeq \nu_{\rm c} \gamma_0^2$, where
$\nu_{\rm c} = eB/2 \pi m_{\rm e} c$) which equals  $\nu_{\rm max}
\sim (10^{13}$-$10^{16})$ Hz for $M = 2 \times 10^8\, \msun$ and
$\nu_{\rm max} \sim (10^{17}$-$10^{20})$ Hz for $M = 10\, \msun$.

We use the synchrotron absorption coefficient (e.g.\  equation 1 in
Ghisellini \& Svensson 1991)  to compute the self-absorption
frequency, $\nu_{\rm t}$ (below which the flow is optically thick to
absorption), in the hybrid electron distribution consisting of the
thermal and the steady-state nonthermal components.  We find (in
agreement with M99) that the presence of pion-decay
electrons has a small effect on $\nu_{\rm t}$; for $M = 10\, \msun$ it
negligibly affects the value of $\nu_{\rm t}$ and for $M = 2 \times
10^8 \msun$ it increases $\nu_{\rm t}$ by a factor of $\sim 2$.  In
all models the Lorentz factor of electrons radiating at $\sim \nu_{\rm
  t}$ is $\gamma \sim 10$ so, with $N(\gamma) \propto \gamma^{-2}$ at
$\gamma \ll \gamma_0$, only a small fraction of the power injected in
pion-decay electrons is emitted below $\nu_{\rm t}$ and most of the
nonthermal emission is available as a seed photon input for Comptonization.

For  $M = 2 \times 10^8\, \msun$, Fig.\ \ref{fig:spec}b compares the
spectra of the synchrotron emission and its thermal Comptonization
observed by a distant  observer in the hadronic and standard models
with the same parameters. As we can see, the much stronger seed photon
flux in the hadronic model gives much lower electron temperatures,
which is reflected in a reduced cut-off energy and much softer X-ray
spectra.  

Fig.\ \ref{fig:spec}c shows the contribution of the synchrotron radiation 
to the {\it observed} spectrum for model HN with $M = 10\, \msun$. Note that due to GR transfer 
effects the observed nonthermal synchrotron radiation has the maximum, in $\nu F_{\nu}$, at 
energies an order of magnitude lower than the rest-frame spectrum.

The high-energy part of the nonthermal synchrotron spectra depends on the
proton distribution. In model HN it is a power-law with the photon spectral index $\Gamma = 1 + s_{\rm p}/2$, as in Fig.\ \ref{fig:spec}c (where, for
$s_{\rm p}=2.6$, $\Gamma=2.3$).
In model HT
it decreases exponentially, as in Fig.\ \ref{fig:spec}a. However, this difference at high-energies
does not affect the impact on electron temperature so, in
particular, the same $T_{\rm e}$ is predicted by model HN and HT with large $a$ (see below).

Fig.\ \ref{fig:synch}a shows the radial distribution of vertically
integrated emissivities of thermal, $Q_{\rm th.s}$, and nonthermal,
$Q_{\rm nth.s}$, synchrotron radiation. Figs \ref{fig:synch}bc show
the ratio of the total nonthermal and thermal synchrotron
emissivities, $Q_{\rm nth.s.tot}$ and $Q_{\rm th.s.tot}$, computed by
integrating $Q_{\rm nth.s}$ and $Q_{\rm th.s}$, respectively, over
$r$.  Open symbols correspond to $Q_{\rm th.s.tot}$ in hadronic (or
DA) models and full symbols to $Q_{\rm th.s.tot}$ in model S; the
difference between them illustrates the amount of the decrease of
$T_{\rm e}$ due to the additional seed photons from nonthermal
synchrotron.  Note that the results of Wardzi\'nski \& Zdziarski
(2001) imply that $Q_{\rm nth.s.tot} / Q_{\rm th.s.tot} \sim 10^5$
should correspond to the supermassive black-hole model, whereas our
model gives much lower values. The difference is due to a small
$kT_{\rm e}=50$ keV assumed in Wardzi\'nski \& Zdziarski and to a very
strong sensitivity of  $Q_{\rm th.s}$ on $T_{\rm e}$, with the increase of
$T_{\rm e}$ by a factor of 2 corresponding to the increase of $Q_{\rm
  th.s}$ by over an order of magnitude. Indeed, if we artificially set
$kT_{\rm e}=50$ keV in our models, we get $Q_{\rm nth.s.tot} / Q_{\rm
  th.s.tot} \sim 10^5$, however, the electron energy balance typically
yields larger $T_{\rm e}$ at $L/L_{\rm Edd} \la 0.01$.

Below we briefly summarize the dependence on $M$, $\dot m$, $a$,
$\delta$ and the proton distribution function.

{\it Heating of electrons}. For large $\delta$, the pion-decay
$e^{\pm}$ have insignificant effect and the hadronic models differ
negligibly from corresponding model S, see stars in
Fig.\ \ref{fig:synch}b. In these models, $L$ is by an order of
magnitude larger than in models with small $\delta$ with  the same
$\dot m$, therefore, the rate of proton-proton interactions (which
depends on density squared, i.e.\ $\propto \dot m^2$) is by two orders
of magnitude lower than in small-$\delta$ models with the same  $L$.
Below we focus on low-$\delta$ models.

{\it Black hole mass}.  We find that in models with $M= 10\, \msun$
the presence of pion-decay $e^{\pm}$ negligibly affects the electron
temperature.  Thermal synchrotron emissivity scales with $M$ as
$Q_{\rm s,th} \sim M^{1/2}$ (Mahadevan 1997). Nonthermal synchrotron
emissivity of $\pi^{\pm}$-decay $e^{\pm}$  scales linearly with $M$
(neglecting insignificant differences due to the
self-absorption). This implies that for a supermassive black hole,
$Q_{\rm s,nth}/Q_{\rm s,th}$ is larger, typically by $\sim 2$-3 orders
of magnitude, than for $M= 10\, \msun$;  see Figs \ref{fig:spec}a and
\ref{fig:synch}abc.  Even at $L \simeq 0.01 L_{\rm Edd}$, when $Q_{\rm
  s,nth,tot}$ exceeds $Q_{\rm s,th,tot}$  for $M= 10\, \msun$ (in
models with small $\delta$),  the  stellar-mass black-hole models
predict higher electron temperatures and harder X-ray spectra than
supermassive black-hole models. This results from the difference of
$\nu_{\rm max}$ noted above;  for $M= 10\, \msun$ the seed photons are
emitted mostly in the X-ray/$\gamma$-ray range and hence the  Compton cooling
is much less efficient.

{\it Accretion rate}.  As seen in Fig.\ \ref{fig:synch}b, the $Q_{\rm
  s,nth,tot}/Q_{\rm s,th,tot}$ ratio strongly increases  with
increasing luminosity, as a result of (1) the  decrease of the
electron temperature with increasing $\dot m$ (due to the increase of
$\tau$) leading to the decrease of $Q_{\rm s,th,tot}$, and (2) the
increase of $Q_{\rm s,nth,tot} \propto \dot m^2$. Then, the difference
between the hadronic and standard model also increases with $L$, see
Fig.\   \ref{fig:spec}b. As also seen in Fig.\ \ref{fig:spec}b, the X-ray
spectra harden with increasing $L$ in both versions of the model. The
hardening is due to a rather slow increase of $\tau$ ($\propto \dot
m$) with increasing $L$. Namely, for heating of electrons dominated by
Coulomb interactions, $L \propto \dot m^\zeta$, with $\zeta = 2.2-2.5$
depending on the model parameters (the increase is slightly faster
than $\dot m^2$ due to the decrease of $T_{\rm e}$) and then $\tau
\propto \dot m^{1/\zeta}$.  The hardening is, however, much stronger
in the standard model in which the flux of seed photons decreases, in
contrast to the hadronic models in which it increases with increasing 
$\dot m$, as noted above.

{\it Black hole spin and proton distribution function}.  The rate of
pion production depends on the number of protons with energies above
the pion production threshold and for the thermal distribution of
protons it strongly depends on the proton temperature, $T_{\rm p}$. In
turn, $T_{\rm p}$ increases with increasing $a$ because the rapid rotation of a
black hole stabilizes the circular motion  of the flow and increases
the dissipation rate; for $a=0$, $T_{\rm p}$ in the innermost part is
typically by a factor of $\sim 4$ smaller than for the maximum value,
$a=0.998$, with other parameters unchanged.  The resulting dependence
of the power emitted by pion-decay electrons on $a$ in model HT is
shown in Fig.\ \ref{fig:synch}d; as we can see, $Q_{\rm
  s,nth,tot}/Q_{\rm s,th,tot}$ differs by three orders of magnitude
between a non-rotating and an extremely-rotating black hole. However,
the dependence on $a$ is significantly reduced in model HN, as also
shown in Fig.\ \ref{fig:synch}d, because for a power-law distribution
of protons the fraction of protons above the threshold is only
linearly dependent on the average proton energy (cf.\ M99). For $a \ga
0.9$, properties of the hadronic model are almost independent of the
proton distribution; the pion production rate for HT and HN is almost the same in the
innermost part where the dominating contribution to $Q_{\rm
  s,nth,tot}$ comes from, see Fig.\ \ref{fig:synch}a.

\begin{figure*} 
\centerline{\includegraphics[height=6.5cm]{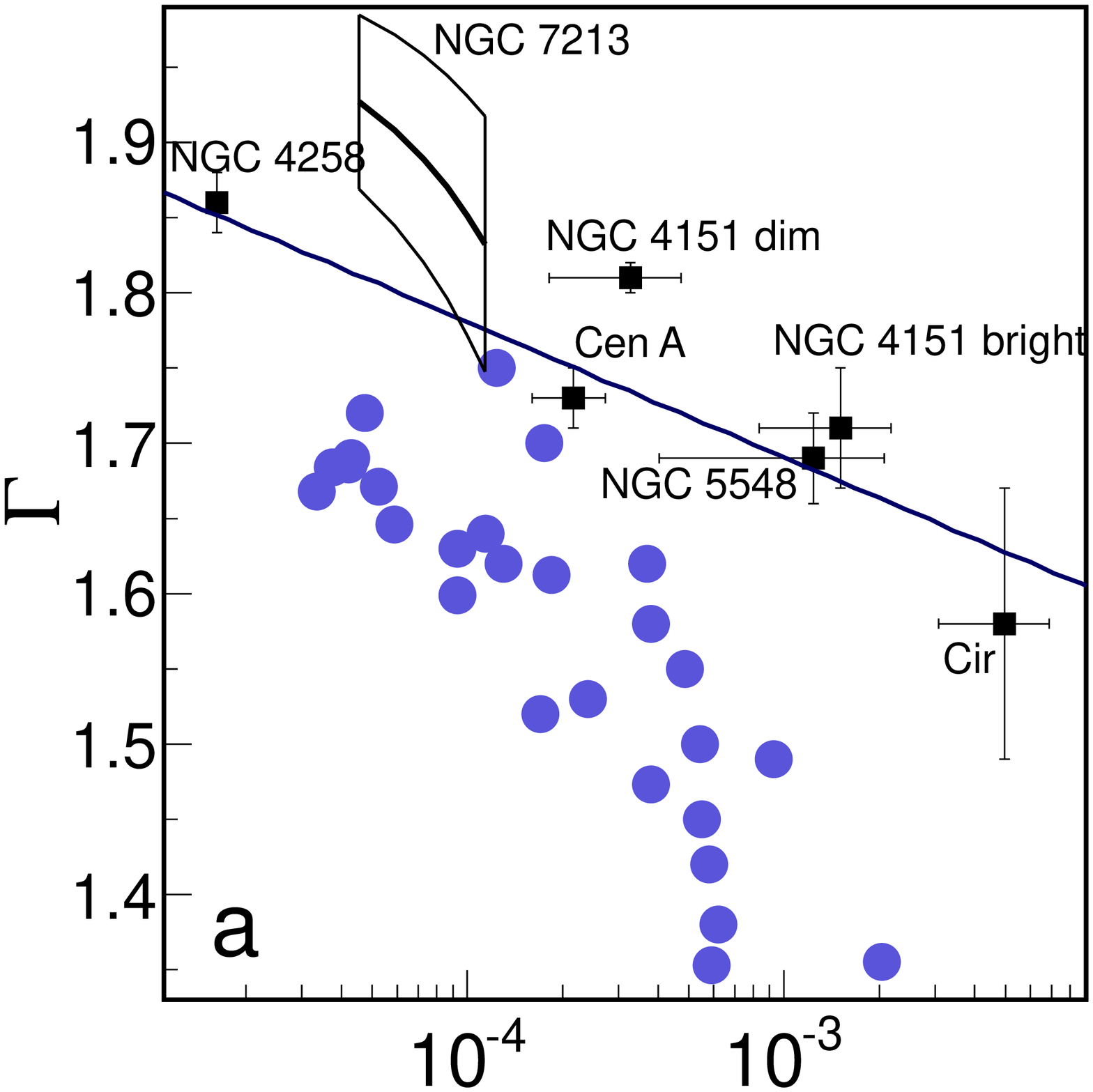}\includegraphics[height=6.5cm]{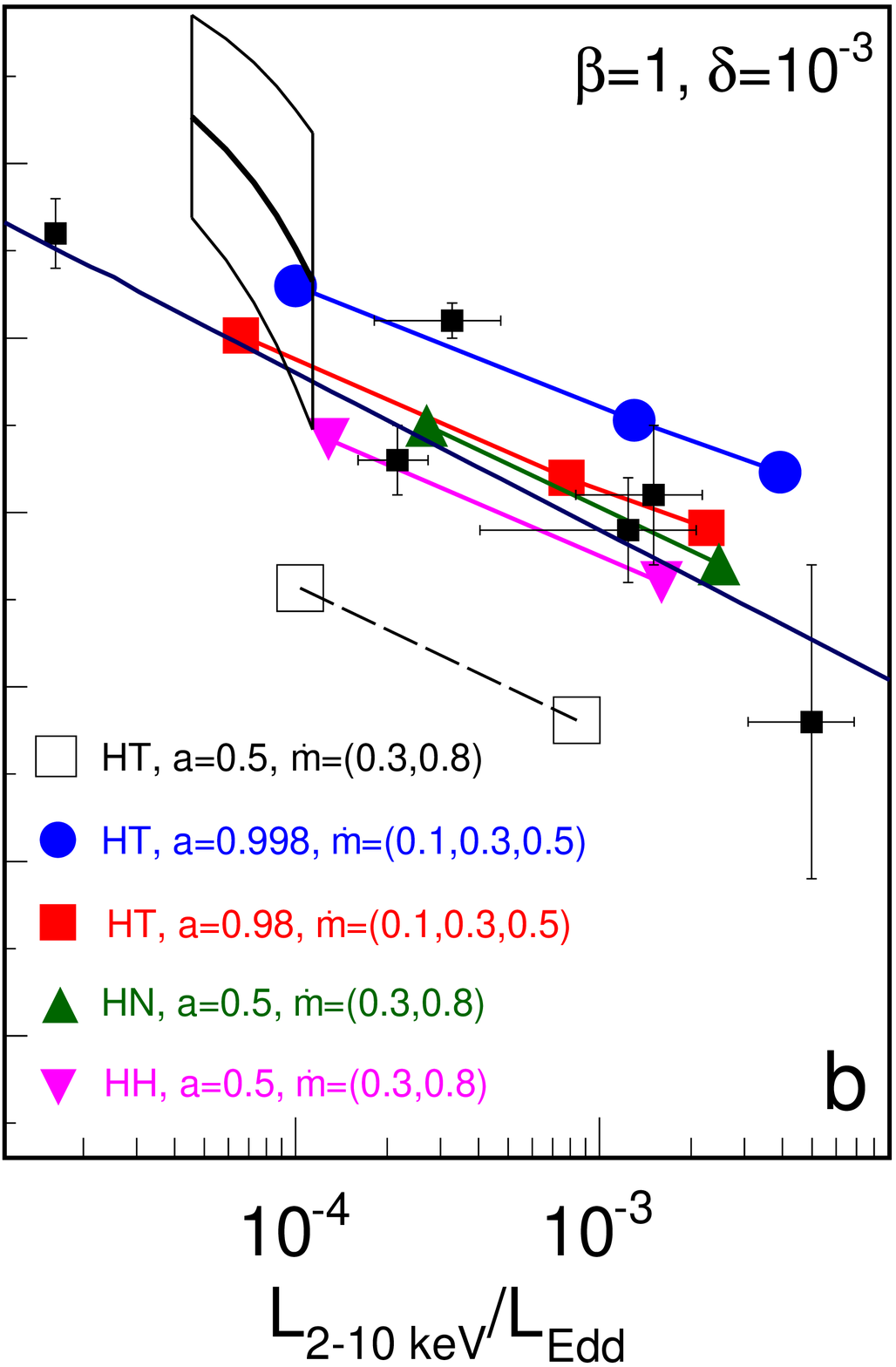}\includegraphics[height=6.5cm]{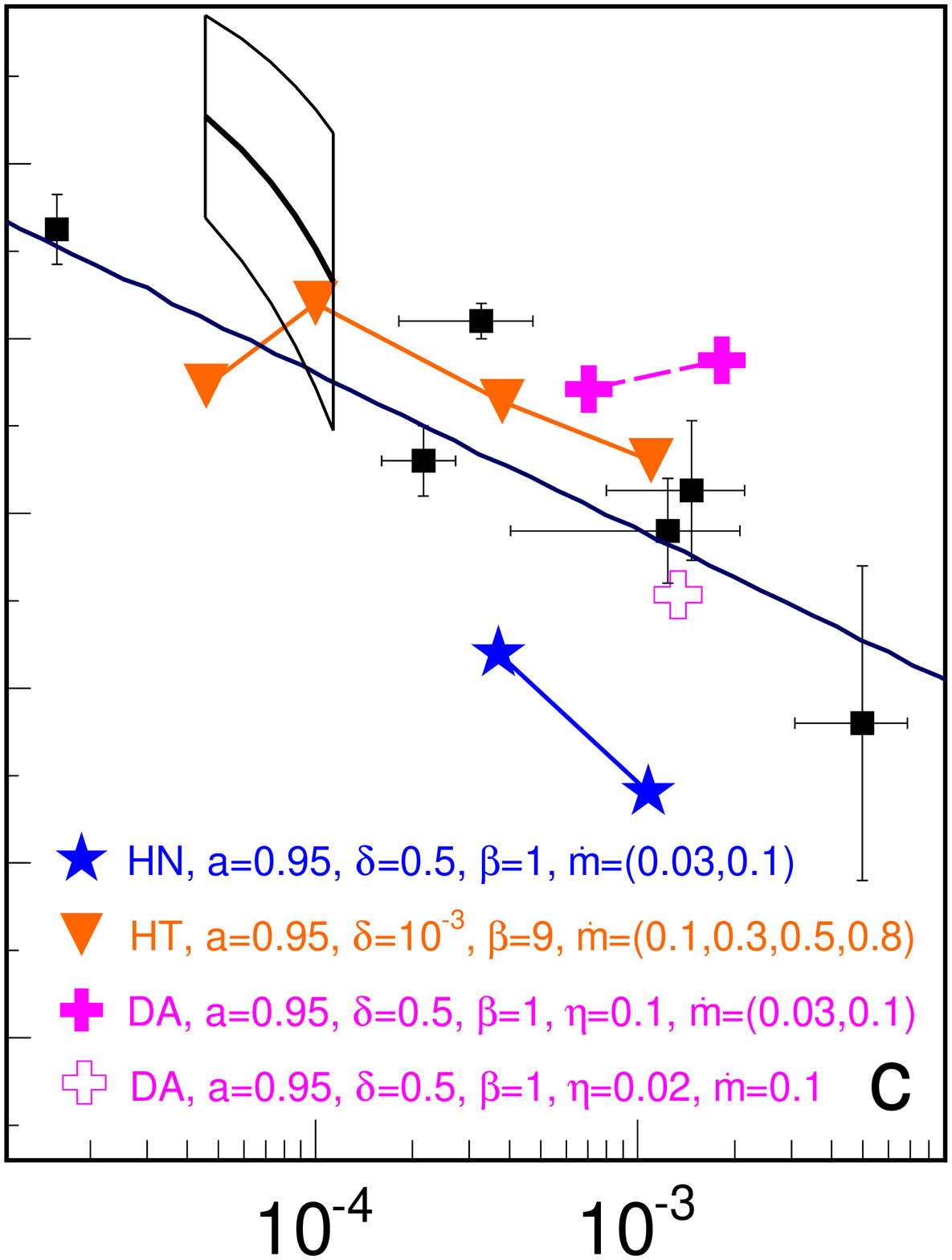}}
\caption{Photon spectral index as a function of the 2-10 keV Eddington
  ratio, predictions of the hot-flow model with $M =2 \times 10^8\, \msun $ are compared with the
  observational data for low-luminosity AGNs (see text; dark blue line
  shows fit from Gu \& Cao 2009). (a) The standard ADAF model S.  (b)
  Hadronic models with $\delta =10^{-3}$ and $\beta = 1$. (c) Hadronic and DA models.}
\label{fig:agn} 
\end{figure*}

\subsection{Directly accelerated electrons}
\label{sect:da}

We consider  model DA for $\delta=0.5$; as noted above, for such large
values of $\delta$ the hadronic effects negligibly affect the electron
temperature. In model DA the importance of self-absorption strongly
depends on details of electron acceleration. If electrons are injected
with a power-law distribution, $n_{\rm inj}(\gamma) \propto
\gamma^{-s_{\rm inj}}$, for $s_{\rm inj} > 2$ most of the injected power is thermalized by
self-absorption  (see e.g.\  Malzac \& Belmont 2009) and
only a small fraction available as the seed photon input. 

To make comparison of  the hadronic model with model DA  not affected
by such differences, we assume an  injection of nonthermal electrons
with $\gamma_0 = 100$ for $M = 2 \times 10^8\, \msun$, similar as for
pion-decay electrons, and $\gamma_0 = 20$ for $M = 10\, \msun$ (this
value is explained below).  For such a monoenergetic injection, the
steady state distribution of electrons is  $N(\gamma) \propto
\gamma^{-2}$ and the self-absorption effects are insignificant, like in
the hadronic models.

 Figs \ref{fig:synch}bc illustrate effects of nonthermal electrons in  model DA with
$\eta=0.1$, which value of $\eta$ gives a similar  $Q_{\rm
  s,nth,tot}/L$ ratio as in model HT with large $a$ or model HN.  We
first note that  for a large (and constant) $\delta$, the $L \propto
\dot m$ scaling implies $\tau \propto L$, i.e.\ $\tau$ increases much faster
with $L$  than for a small $\delta$. As a result, $T_{\rm e}$
decreases with increasing $L$ faster in large-$\delta$ models, leading
to a stronger decrease of $Q_{\rm th.s.tot}$. In model S (with large $\delta$), it results in
hardening of the X-ray spectra with increasing $L$.  In model DA with
$M = 2 \times 10^8\, \msun$, however, the seed photon flux increases
proportionally to $L$ 
and this, combined with the fast increase of $\tau$, leads to a softening of the
X-ray spectra (illustrated in Fig.\ \ref{fig:agn}c below). 

For $M = 10\, \msun$, the presence of nonthermal electrons in model DA
with $\eta=0.1$ negligibly affects the electron temperature.  Here,
similarly as in hadronic models, $Q_{\rm nth.s.tot}$  exceeds $Q_{\rm
  th.s.tot}$ only at $L \ga 0.01 L_{\rm Edd}$.  To check the maximum
possible impact of the nonthermal synchrotron radiation at $L \sim
0.01 L_{\rm Edd}$, we  have assumed $\gamma_0 \sim 20$ for which most
of the nonthermal synchrotron photons are emitted at $\sim 1$ keV
(i.e.\ at lower energies than in the hadronic model), not much higher
than $h \nu_{\rm t}$. Even for such $\gamma_0$, the Compton  cooling
rate in Eddington units is by a factor of several  smaller than in the
same model DA with   $M = 2 \times 10^8 \msun$; see also Dermer et
al.\ (1991) for the dependence of the cooling rate on the seed photon
energy.

\begin{figure*} 
\centerline{\includegraphics[height=8cm]{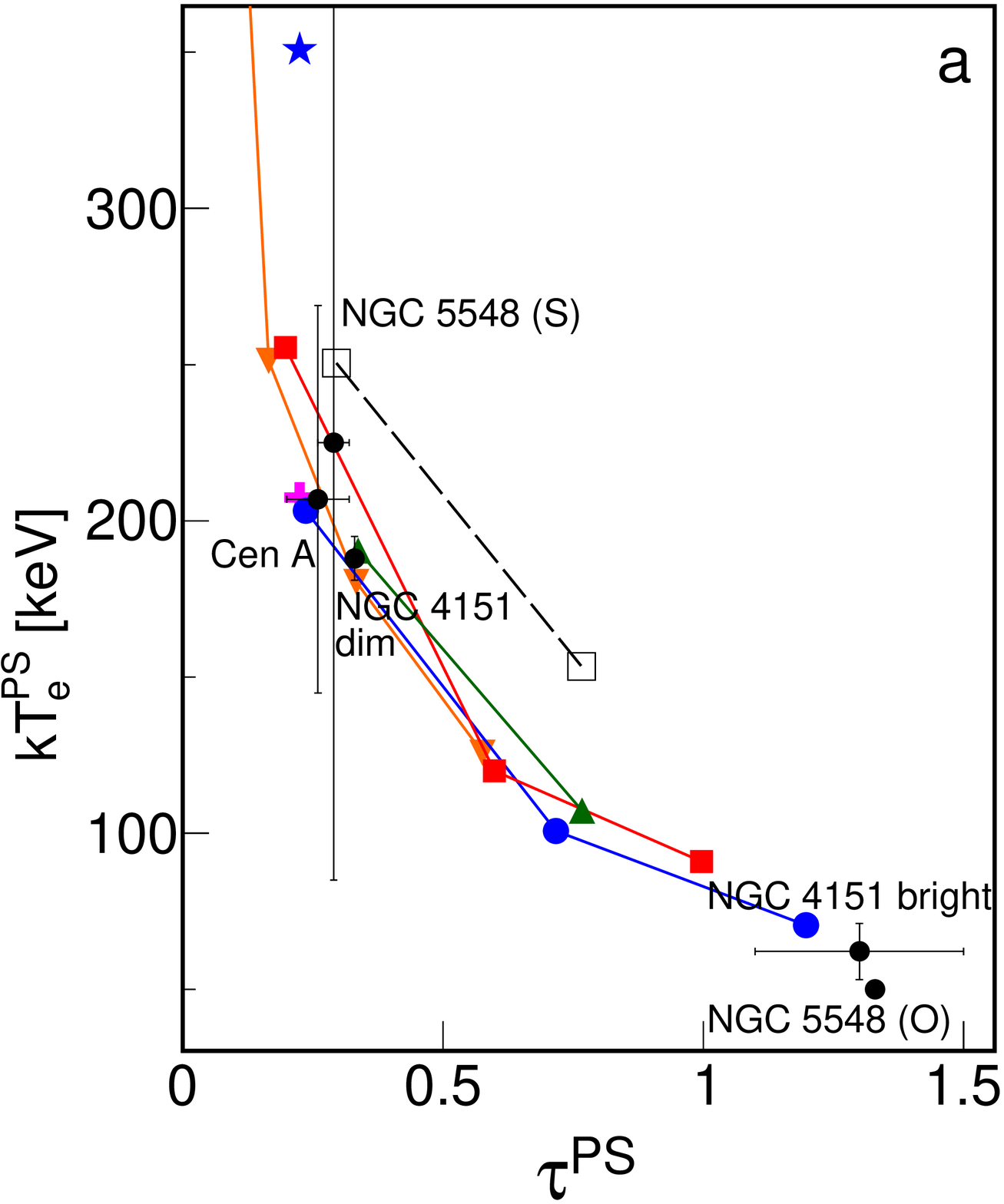} \includegraphics[height=8cm]{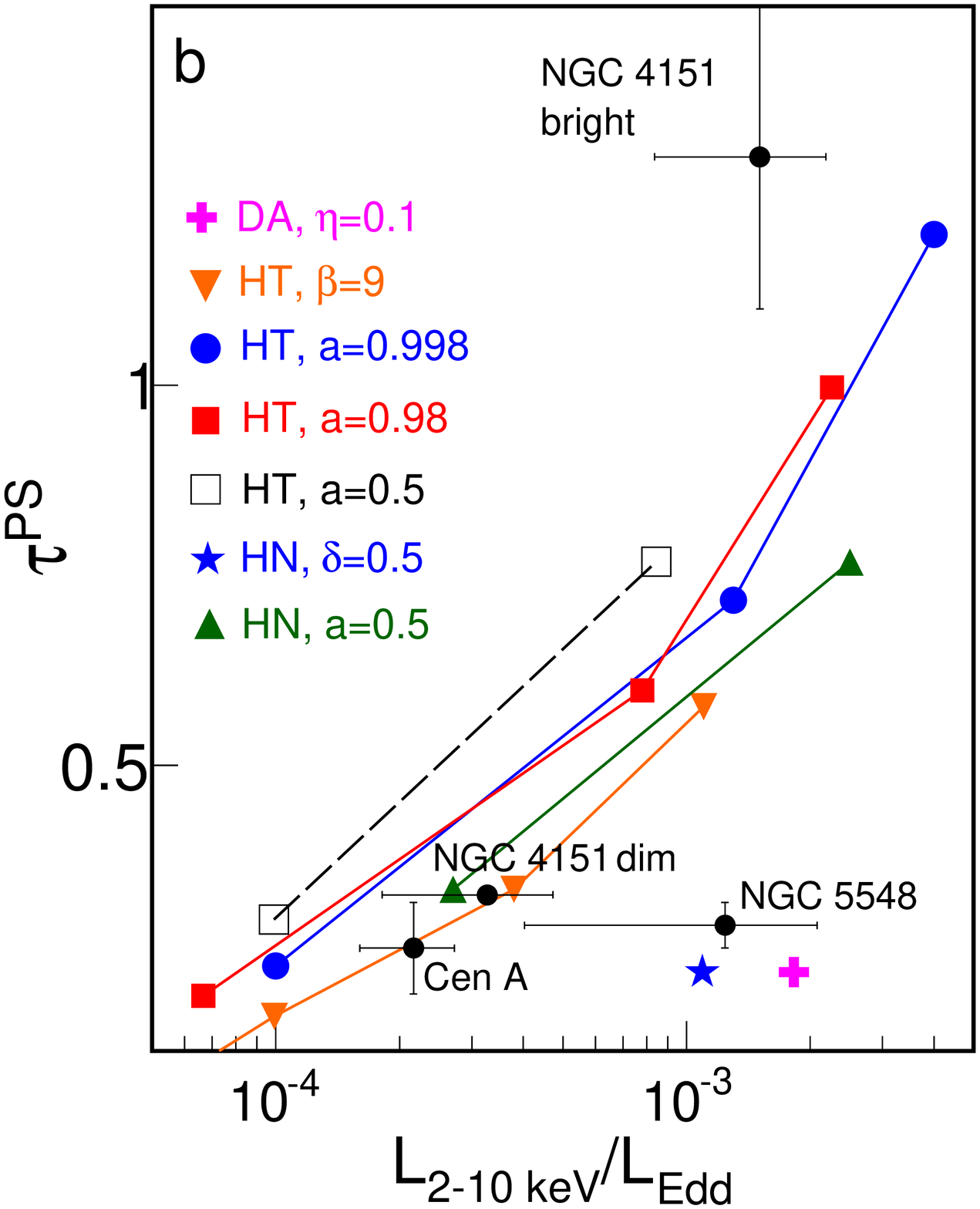}}
\caption{(a) $T_{\rm e}^{\rm PS}$ as a function of $\tau^{\rm PS}$,
  (b) $\tau^{\rm PS}$ as a function of $\lambda_{2-10}$;  parameters
  of the COMPPS slab model  best-matching the hot-flow spectra
  compared with plasma parameters measured in several AGNs (see
  text). In both panels:  (blue) circles are for model HT with
  $a=0.998$, $\delta =10^{-3}$, $\beta = 1$, $\dot m=0.1$, 0.3 and
  0.5; (red) squares are for model HT with  $a=0.98$, $\delta =10^{-3}$, $\beta = 1$, $\dot m=0.1$,
  0.3 and 0.5; (green) triangles-up are for model HN with $a=0.5$, $\delta =10^{-3}$, $\beta = 1$,
  $\dot m=0.3$ and 0.8; (black) open squares are for model HT with $a=0.5$, $\delta =10^{-3}$, $\beta = 1$,
  $\dot m=0.3$ and 0.8; (orange) triangles-down  are for model HT
  with  $a=0.95$, $\beta=9$, $\delta =10^{-3}$, $\dot m=0.1$,
  0.3, 0.5 and 0.8; (blue) star is for model HN with  $\beta=1$,
  $\delta =0.5$,  $\dot m=0.1$; (magenta) cross is for model DA
  with $a=0.95$, $\delta=0.5$,  $\dot m=0.1$ and $\eta=0.1$.
}
\label{fig:tetau} 
\end{figure*}

\section{Comparison with low-luminosity AGNs}

In Fig.\ \ref{fig:agn} we compare the $\Gamma$--$\lambda_{2-10}$
(where $\lambda_{2-10} = L_{2-10}/L_{\rm Edd}$) relation predicted in our
model  with the observational data for
low-luminosity AGNs, same as used in N14.  The sample includes nearby
AGNs with precisely measured black-hole masses and the spectra
measured with high-quality statistics, which allows to estimate the
intrinsic X-ray spectrum. The values of the intrinsic $\Gamma$ and
$L_{2-10}$ are taken from  Lubi\'nski et al.\ (2010) for the dim and
bright states of NGC 4151, Brenneman et al.\ (2012) for NGC 5548,
Beckmann et al.\ (2011) for Centaurus A, Yamada et al.\ (2009) for NGC
4258 and Yang et al.\ (2009) for Circinus; see N14 for details.  To
find $\lambda_{2-10}$ we use black-hole mass measurements quoted in
N14, except for NGC 4151 for which we use the new stellar dynamical
mass measurement, $M = (3.76 \pm 1.15) \times 10^7\, \msun$ (Onken et
al.\ 2014).  We also use fits to the $\Gamma$--$\lambda_{2-10}$
anticorrelation in a large sample of low-luminosity AGNs from Gu \& Cao (2009) 
and in NGC 7213 from Emmanoulopoulos et al.\ (2012).

Fig.\ \ref{fig:agn}a shows the model points for model S, including all
results for models with $M =2 \times 10^8\, \msun $ from N14 and
several additional solutions for model S corresponding to hadronic
models computed in this work.  The clear discrepancy between the model
and the data was noted in Section 1 as a motivation for this work.
The two model points located close to Cen A are for the models (with
$\delta=0.5$) with very tenuous flows, having $\tau \la 0.01$ and
$kT_{\rm e} \ga 1$ MeV; extension of this model to lower $L$ can
approximately reproduce the relation assessed in NGC 7213 if the
average slope of the X-ray spectrum is considered, however, these
X-ray spectra strongly deviate from a power-law.

As we can see in Fig.\ \ref{fig:agn}b, the hadronic model with small
$\delta$ reproduces the $\Gamma$--$\lambda_{2-10}$ relation observed
in AGNs.  Model HN agrees with the AGN data regardless of the
black-hole spin value; note also that model HH gives a similar
$\Gamma$--$\lambda_{2-10}$ relation to model HN, so it is not very sensitive 
to the fraction of total energy in the nonthermal proton
component. Model HT agrees with the data for large values of $a$; for
$a \la 0.9$ it predicts slightly too hard spectra, e.g.\ by $\Delta
\Gamma \simeq -0.1$ for $a=0.5$.

The model solutions shown in Fig.\ \ref{fig:agn}b correspond to
$\beta=1$. As we see in Fig.\ \ref{fig:agn}c, the hadronic model with
$\beta=9$ gives a similar relation, except for the lowest
$\lambda_{2-10}$ for which the compressive heating of electrons
dominates, leading to $\tau \propto L$, and then in this range we get
a positive correlation (similarly as in model DA, see below). Stars in
Fig.\ \ref{fig:agn}c are for the hadronic model with $\delta=0.5$, which differs 
only weakly from model S as discussed in  Section \ref{sect:pion}.

Dashed line in Fig.\ \ref{fig:agn}c  illustrates the positive $\Gamma$--$\lambda_{2-10}$
correlation in model DA with a constant $\eta=0.1$ which, as noted in
Section \ref{sect:da}, results from (1)  the linear increase of $\tau$ with
$L$, combined with (2) the linear increase of seed photon flux with
$L$. The model with $\eta \sim 0.1$ can explain the intrinsic X-ray slopes
measured in AGNs. However, the decrease of $\eta$ is required (if
$\delta$ does not decrease significantly with increasing $\dot m$), by
a factor of $\sim 2$ for the increase of $L$ by a factor of 3, to
reproduce the observed negative relation.

Finally, we note that the slope of  the $\Gamma$--$\lambda_{2-10}$
correlation observed in individual AGNs, NGC 4151 (the change between
dim and bright states) and NGC 7213, is slightly steeper than
predicted by the hadronic models. Then, an additional, weak
contribution of directly accelerated electrons may be required at
lower $\dot m$.

Further constraints on physics of hot flows can be obtained by modeling
of the observed spectra with thermal Comptonization model. It allows
to estimate the electron temperature, which is directly related to the
cut-off energy, and the optical depth, which in combination the
temperature determines the slope of the spectrum.  However, precise
measurements of the high-energy cut off are necessary for such
modeling, which is a challenging task for AGNs due to poor quality of
the hard X-ray data.  Such measurements are available only for a few
brightest AGNs, furthermore, there is a discrepancy in results
based on data provided by different detectors (see
e.g.\ discussion in Lubi\'nski et al.\ 2010).

Similarly as in N14, we have fitted the hot flow spectra 
using the  slab COMPPS model (Poutanen \& Svensson
1998).  The fitted parameters, $T_{\rm e}^{\rm PS}$ and $\tau^{\rm
  PS}$, are compared  in Fig.\ \ref{fig:tetau} with plasma parameters
assessed in AGNs with relevant $\lambda_{2-10}$. We use the COMPPS fits for NGC 4151 (Lubi\'nski et
al.\ 2010), Cen A (Beckmann et al.\ 2011) and NGC 5548 (Brenneman et
al.\ 2012). In Fig.\ \ref{fig:tetau}a we plot also the thermal
Comptonization fit for NGC 5548 from Magdziarz et al.\ (1998). The
parameters from Brenneman et al.\ (2012) and Magdziarz et al.\ (1998)
are denoted  by S ({\it Suzaku}) and O (OSSE), respectively; their
fits used the sphere models, so we use the fitted $\tau$ reduced by a
factor of 2 and 1.5, respectively, which reduction factors correspond
to the fitted values of $\tau$ (see  N14).

As we see, various hadronic models with small $\delta$  (except for
model HT with small or moderate $a$) predict similar $T_{\rm e}^{\rm
  PS}(\tau^{\rm PS})$ relations, which are in agreement with plasma
parameters measured in AGNs.  Similarly, model DA with $\eta = 0.1$
agrees with the AGN data at  $\tau^{\rm PS} \simeq 0.2$.  In N14 we
found that lower values of $\beta$ give lower $T_{\rm e}^{\rm PS}$
through the change of the flow geometry. Namely, lower-$\beta$ flows
are closer to a slab and higher-$\beta$ flows are closer to a sphere
and, thus, the Compton cooling is less efficient in the latter.  In the
hadronic models, however, the dependence of the  $T_{\rm e}^{\rm
  PS}(\tau^{\rm PS})$ relation on $\beta$ is reduced; the
large-$\beta$ flows have much larger proton temperatures and the
larger pion-production rate approximately compensates the difference
in geometry.

Larger differences between the model predictions, especially those for
high and low values of $\delta$, occur in the $\tau^{\rm
  PS}(\lambda_{2-10})$ relation (Fig.\ \ref{fig:tetau}b).   At
$\lambda_{2-10} \sim 10^{-3}$, $\delta=0.5$ gives $\tau^{\rm PS}$ by a
factor of several lower than $\delta=10^{-3}$.  As we see, values of $\tau^{\rm PS}$ estimated in
NGC 4151 and Cen A strongly favor low values of $\delta$. In turn,
$\tau^{\rm PS}$ estimated from the {\it Suzaku} data of  NGC 5548 is
consistent with large $\delta$.  However, we emphasize again that the
estimation of $\tau^{\rm PS}$ strongly relies on the precise
determination of the spectral cut-off and the discrepancies between  different
studies of the observed data, noted above, are similar in magnitude to
the differences in the model predictions.  The observational
uncertainty is illustrated by the difference between plasma parameters
for NGC 5548 obtained using the {\it Suzaku} and OSSE data; almost the
same difference for the bright state of NGC 4151 follows from studies
of the {\it Integral} data  by Lubi\'nski et al.\ (2010) and  {\it
  BeppoSAX} data by Petrucci et al.\ (2001).

\section{Summary and discussion}

The $\Gamma$--$\lambda_{2-10}$ relation, which can be robustly
determined in nearby AGNs, indicates that a source of seed photons
much stronger than thermal synchrotron is needed to explain the X-ray
radiation of low-luminosity AGNs by thermal Comptonization in hot
accretion flows. Using a precise hot-flow model we find that the
nonthermal synchrotron emission of relativistic electrons is a good
candidate for a sufficiently efficient source.  If the power in the
nonthermal electron component equals $\sim 10$ per cent of the power
used to heat the thermal electrons, the model can explain the observed
relation essentially for any value of $a$, $\beta$ or $\delta$.  The
origin of the non-thermal component with such a magnitude can be
explained by two different mechanisms and it depends on the value of
$\delta$.

 If thermal electrons are heated mainly by Coulomb interactions (low
 $\delta$), the nonthermal electrons may come from the decay of pions
 produced by proton-proton interactions. The flux of nonthermal
 synchrotron photons is then determined by the hydrodynamic solution
 and the proton distribution function. Remarkably, the hadronic
 collisions produce a sufficient amount of nonthermal electrons for a
 broad range of parameters, excluding only a moderately or slowly
 rotating black hole with a thermal distribution of protons.  The idea
 that relativistic protons may be responsible for the production of
 relativistic electrons was considered in several works in 1980s
 (e.g.\ Kazanas \& Ellison 1986, Zdziarski 1986, Sikora et al.\ 1987),
 following realization that the direct Fermi acceleration of electrons
 is much less efficient than the acceleration of protons and, thus,
 electrons produced by pion decay should outnumber those directly
 accelerated.

If a large fraction of  the accretion power is used for the direct
heating of thermal electrons (large $\delta$), the nonthermal
component must be due to direct acceleration of electrons. Such
acceleration of electrons is indeed possible in hot flows according to
recent MHD simulations (e.g.\ Ding et al.\ 2010, Riquelme et al.\ 2012). 
In this version of the model, a specific scaling of the thermal heating and acceleration
parameters, $\delta$ and $\eta$, with the accretion rate is needed to explain the observed
$\Gamma$--$\lambda_{2-10}$ anticorrelation. We note that  although $\delta$ and $\eta$
are treated as independent parameters in our computations, they are certainly related
to the same microphysical processes that convert the accretion power into the plasma
kinetic energy. We also emphasize that a value of $\eta \sim 0.1$,
sufficient to explain the observed data, was assessed under the assumption of monoenergetic 
injection of electrons. A more realistic scenario, with electrons accelerated into a power-law
distribution with the index $s_{\rm inj}$, is properly approximated by the monoenergetic injection
only for $s_{\rm inj}<2$. For larger $s_{\rm inj}$ (which are supported by some observations, see below), a significant fraction of nonthermal energy is self-absorbed
(for $s_{\rm inj} \ga 3.5$ this may be the only form of thermal plasma heating, cf.\ Malzac \& Belmont 2009)
and then a larger value of $\eta$ would be required. 

The plasma parameters predicted by the large-$\delta$ and
small-$\delta$ models differ significantly and a precise estimation of
the electron temperature and optical depth should allow to constrain
the value of $\delta$.  However, results provided by different
detectors disagree even for brightest AGNs. Results of some studies
(based on OSSE or {\it Integral} data) seem to favor low values of
$\delta$, while other (using {\it BeppoSAX} or {\it Suzaku} data)
appear more consistent with large $\delta$.

In flows surrounding stellar-mass black holes, the nonthermal
synchrotron is emitted in the X-ray/$\gamma$-ray range. Therefore, the Compton
cooling is much weaker, and the presence of the nonthermal electron
component receiving $\sim 10$ per cent of the total power (due to
pion-decay for small $\delta$ or direct acceleration for large
$\delta$, similarly is in AGN models) insignificantly affects the
electron temperature.  The X-ray spectra predicted by both the
hadronic and direct acceleration models for $M = 10\, \msun$ differ
only weakly from the spectra predicted by the standard model (with the
thermal synchrotron only), which were studied in detail in N14. In N14
we found that the evolution of black-hole transients   agrees with
predictions of the standard model, in particular they reach $\Gamma
\simeq 1.4$ at $L \sim 0.01 L_{\rm Edd}$ (as compared to the intrinsic
$\Gamma \ga 1.7$ measured in AGNs at such $L$).  The results of this
work indicate that the  agreement does not exclude the presence of a
significant nonthermal electron component.

 We conclude that the hot flow model with seed photons produced by the
 nonthermal synchrotron radiation provides an attractive description
 for all accreting black-hole systems, with the difference between
 AGNs and black-hole transients around $\sim 0.01 L_{\rm Edd}$
 explained by the scaling of basic physical processes with the
 black-hole mass.

Finally, we comment on the observability of spectral  features formed by pion-decay $e^{\pm}$. 
The annihilation feature is not observable, e.g.\ for the model shown in Fig.\ \ref{fig:spec}(c)
the observed luminosity of the annihilation photons is $\sim 10^{33}$ erg/s.
The high-energy part of the synchrotron radiation perhaps could be observed in bright nearby objects, like Cyg X-1.
However, the MeV tail which is observed in Cyg X-1 at $L \simeq 0.01 L_{\rm Edd}$  (e.g.\ Del Santo et al.\ 2013) is over an order of magnitude stronger than the synchrotron component predicted by the model, as shown 
in Fig.\ \ref{fig:spec}(c) for the solution with  $L \simeq 0.007 L_{\rm Edd}$. 
If the observed tail is produced in the hot flow,
it indicates that an efficient direct acceleration must take place in the flow, 
because the $\pi^{\pm}$-decay electrons cannot produce a tail with such a magnitude at such $L$. However, the tail may be produced in other sites, e.g.\ in a cold disc corona, as suggested 
e.g.\ by the increase of the relative strength of the nonthermal component 
with the increasing intensity of the thermal disc emission, assessed by Del Santo et al.\ (2013).
We note also that the high-energy tail in GX 339-4 was revealed in the rising phase of GX 339-4 by  Droulans et al.\ (2010) at $L/L_{\rm Edd}$ much larger than these of our solutions and the results of this work cannot be used to estimate the corresponding strength of pion-decay component. A relevant hot flow model,
with strong Coulomb cooling of protons, would be characterized by
a much larger density and hence, for a nonthermal proton distribution, by an enhanced pion production.

Interestingly, 
spectral modeling of the above observations of nonthermal tails with hybrid models indicate that a steep injection index of electrons, $s_{\rm inj} \ga 2.5$, is required. Such a steep injection can be explained by the decay of pions for a steep acceleration index of protons, such as assumed in our computations.

\acknowledgments

We thank the referee for useful comments.
This research has been supported in part by the Polish NCN grant N N203 582240. FGX is supported by the National Basic Research Program of China (973 Program, grant 2014CB845800), the NSFC (grants 11203057, 11103061, 11133005 and 11121062), and the Strategic Priority Research Program "The Emergence of Cosmological Structures" of the Chinese Academy of Sciences (Grant XDB09000000).

\end{document}